\documentstyle[12pt]{article}
\begin{document}
\baselineskip=.285in

\def\be{\begin{equation}}
\def\ee{\end{equation}}
\def\bea{\begin{eqnarray}}
\def\eea{\end{eqnarray}}
\def\ba{\begin{array}}
\def\ea{\end{array}}
\def\bce{\begin{center}}
\def\ece{\end{center}}

%\preprint{DPNU-95-xx}

\title{\Large\bf Solitonic Island in the Bubble
\protect\\[0mm]\ }
\author{\normalsize Yoonbai Kim${}^{(1)}$, Kei-ichi Maeda${}^{(2)}$
and Nobuyuki Sakai${}^{(2)}$\\
{\normalsize\it ${}^{(1)}$Department of Physics, Pusan National University,
Pusan 609-735, Korea}\\
{\normalsize\it ${}^{(2)}$Department of Physics, Waseda University,
Shinjuku-ku, Tokyo 169-50, Japan}}
\date{}
\maketitle

\begin{abstract}
We review how the continuous symmetry can support a soliton inside 
a high-temperature bubble at the time of its nucleation. This solitonic 
island in disoriented phase remains stable during the growth of bubbles
before their collision.
\end{abstract}

\vspace{5mm}

There exists a well-established semiclassical theory of the first-order
phase transition of which the process is described by the formation
and the growth of bubbles \cite{Lan,Col}. The scenario claimed by
this theory is summarized as follows: In order to understand the basic
features of first-order phase transition, it is sufficient 
to consider a model of a real scalar field of which the scalar potential
has a false vacuum and a true vacuum. Then there is unique decay
channel described by a well-known (Euclidean) bounce solution and
the inside of the bubble does not contain any matter lump at the
moment of its nucleation. Though the
above results have been obtained in a specific series of models of 
a real scalar field \cite{Col,Aff}, it is widely believed that, when
the bubble is nucleated, such results may not be changed much for 
more general cases, {\it e.g.} the inclusion of continuous symmetries
or gauge fields \cite{CGM}, but not the gravitational field
\cite{CL}. When we consider the first-order phase transition by
thermal fluctuation, the similar formalism based on the 
$O(3)$-symmetric bubble solution can also be applied \cite{Aff}.

Here several questions may be arisen in accord with the principles of
gauge theories: 1. Why does the continuous symmetry play no important
role at the time of bubble nucleation? 2. Why is the bubble solution
unique even for a variety of models, which depicts only one possible
decay mode from a metastable state to the true vacuum? 3. Can a
matter aggregate like soliton be formed inside a bubble when the 
bubble is nucleated?

In this note we shall address the above questions by considering a
field theory model at finite temperature both in flat spacetime \cite{Kim}
and in curved spacetime \cite{KMS}. Key ingredient to resolve those
questions is to introduce an appropriate continuous symmetry. 
Specific model of our interest is composed of an isovector
$\phi^{a}$ ($a=1,2,3$) with global $O(3)$ symmetry described by the
(Euclidean) action 
\be
S_{E}=\int^{1/T}_{0}\!\!d\tau\int\!d^{3}x\,\biggl\{\frac{1}{2}g^{\mu\nu}
\partial_{\mu}\phi^{a}\partial_{\nu}\phi^{a}+V\biggr\},
\ee
where the Einstein-Hilbert action term and the Jacobian for spacetime
volume should be added in the case of curved spacetime. Any scalar 
potential with one false vacuum $\phi=\phi_{-}$ and one true vacuum
$\phi=\phi_{+}$ among which one is 1 symmetric and
the other is degenerate makes no big difference though a sixth-order 
scalar potential with a false symmetric vacuum at $\phi(\equiv
\sqrt{\phi^{a}\phi^{a}})=0$ and true broken vacua at $\phi=v$ will be
used for actual numerical computation. 

In high temperature limit ($T\rightarrow\infty$) time-dependence 
is neglected and then the contribution of $O(3)$ symmetric bubble,
{\it i.e.}, $\phi=\phi(r)$, 
dominates. A natural choice of scalar phase $\hat{\phi}^{a}=
\phi^{a}/\phi$ is trivial one $(\hat{\phi}^{a}=(0,0,1)$ which
constitutes the well-known $O(3)$ symmetric bubble solution 
\cite{Aff} of minimum action \cite{CGM} (we call it ``{\it normal}''
bubble from now on). Another possibility worth tackling when the
theory of interest contains $O(3)$ symmetry is to take the hedgehog
ansatz $\hat{\phi}^{a}=\hat{r}^{a}$ and then to explore whether
the system really supports the  different bubble solution satisfying
the boundary condition that scalar amplitude should have the false 
vacuum value at spatial infinity. It can rigorously be proved in
flat spacetime that there always exists the bubble solution under
hedgehog ansatz wherever the system contains {\it normal} bubble
\cite{KKK} and a numerical solution of scalar field are given in Fig.
1 (we name this new bubble solution ``{\it solitonic}'' bubble).

%\input{abfig1}

% GNUPLOT: LaTeX picture
\setlength{\unitlength}{0.240900pt}
\ifx\plotpoint\undefined\newsavebox{\plotpoint}\fi
\sbox{\plotpoint}{\rule[-0.200pt]{0.400pt}{0.400pt}}%


\vspace{20mm}

Since the boundary value at the origin should be fixed to the  
false vacuum, {\it i.e.} $\phi(r=0)=\phi_{-}(=0)$, there remains a 
false-vacuum core inside
the true-vacuum region of the bubble due to the winding between
internal $O(3)$ symmetry and spatial rotation. The size of this matter
island is determined by that of the inner bubble wall of order 
$R_{in}\sim 1/m_{meson}$
and the long-range tail of energy density near $\phi\approx\phi_{+}(=v)$, 
{\it i.e.} $T^{t}_{\;t}\sim v^{2}/r^{2}$ in thin-wall cases
($R_{in}\ll r< R_{out}^{n=1}$),
characterizes it as a global monopole formed at the center of
``{\it solitonic}'' bubble (see Fig. 2). Even though there is no-go theorem 
(Derrick Theorem) for finte-energy static solitons composed of scalar 
matters in spacetime 0 more than two, the formation of 
global monopole with the cutoff by the size of {\it solitonic} bubble 
wall is not affected by it since the amount of energy to sustain the tail
of global monopole is proportional to the bubble radius 
($\sim R_{out}^{n=1}$) however the energy proportional to the volume 
of bubble ($\sim R^{n=1^{3}}_{out}$) comes into the bubble as the bubble 
radius increases. 
Fig. 2 also shows that the radius of {\it solitonic} bubble,
$R^{n=1}_{out}$, is larger
than that of {\it normal} bubble, $R_{out}^{n=0}$. It can easily be 
understood by the 
conservation of energy, {\it i.e.} the additional energy used to make a 
matter aggregate is equal to
the loss of energy due to the increase of the radius of bubble.

The effect of monopole island inside the bubble is also drastic in curved
spacetime. Under the Planck scale, the long-range tail of global
monopole renders the spacetime region between the inner and outer walls
flat with the deficit (solid) angle $\Delta=8\pi^{2}Gv^{2}$ and then an
observer must notice the light bending due to the angular separation
$\delta\varphi\sim 8\pi^{2}Gv^{2}\sim {\rm few}\; arcsec$ for the typical
grand unified scale \cite{BV}. It has been known that the global 
monopole inside the {\it solitonic} bubble does not constitute a 
nonabelian black hole even in Planck scale \cite{HL}, however the issues 
such as
the topological inflation at the soliton core \cite{Vil} or the evolution of 
wormhole \cite{MSSK} are intriguing.

\vspace{10mm}
%\input{abfig2}

% GNUPLOT: LaTeX picture
\setlength{\unitlength}{0.240900pt}
\ifx\plotpoint\undefined\newsavebox{\plotpoint}\fi
\sbox{\plotpoint}{\rule[-0.200pt]{0.400pt}{0.400pt}}%


\vspace{20mm}

Even though we found another decay channel described by {\it 0}
bubble, this channel may be physically insignificant if its nucleation 
rate is negligibly small.
The decay probability per unit time per unit volume
is given by exponential form $\Gamma\sim Ae^{-B}$, where $B$ is replaced by
a value of Euclidean action for a bubble solution and $A$ is
estimated by integrating out the fluctuations around a given bubble
solution. It is extremely hard to compute $A$ exactly, however an estimate
based on the zero modes of the fluctuations was suggested in the last paper 
of Ref.\cite{Aff}. Therefore the ratio of decay rates
for both {\it normal} and {\it solitonic} bubbles is expressed by
\be
\frac{\Gamma_{sol}}{\Gamma_{nor}}\sim\biggl(\frac{\tilde{S}_{sol}}{
\tilde{S}_{nor}}\biggr)^{\frac{6}{2}}\exp\Bigl[-\frac{v}{T}(\tilde{S}_{sol}
-\tilde{S}_{nor})\Bigr],
\ee
where $\tilde{S}$ is dimensionless action rescaled by the vacuum
expectation value $v$ and the number of zero modes in this system is six 
among which three
are due to spatial translations and other three are due to rotations.
When $v/T\sim 10^{-1}$, the order of above ratio is $10^{-1}$ in a thin-wall
case and it is around 10 in a thick-wall case under sixth-order
scalar potentials. This tells us a possibility that, though the value of
action for the {\it solitonic} bubble is always larger than that of
{\it normal} bubble, there may exist some region of scalar potential that
the {\it solitonic} bubble becomes more likely to be nucleated.

Once a bubble is formed, its time evolution is of interest. In flat spacetime
case we should take into account thermal effect which can be described
by the physics of combustion process when the environment keeps the
temperature high enough \cite{Ste}. When one considers it in early
universe, the background universe is rapidly cooled down due to gravitational
effect
and the motion of bubbles may follow the classical dynamics because of
the recovery of real Minkowski time. Here we suppose the case 
in curved spacetime that 
the evolution of bubbles is governed by time-dependent field equations
and the other effects are included in the change of initial bubble
shapes, which was indeed the case for the ``{\it normal}'' bubbles in early
universe. Numerical analysis can be summarized in the following as shown
in Fig. 3. ($H$ in Fig. 3 is the Hubble parameter defined by
$H=\sqrt{\frac{8\pi G}{3}V(\phi=0)}$ ) Since the static solution 
depicts the bubble of critical size,
it starts to grow when the initial size of bubble is larger than the 
critical size however smaller one shrinks. Therefore the outer wall of 
``{\it solitonic}'' bubble also expands and its velocity reaches a terminal
value which is 
the same as that of ``{\it normal}'' bubble since the effect of global
monopole formed in the ``{\it solitonic}'' bubble is negligible for
large thin-wall bubbles 
and is smaller than the light speed due to gravitational effect \cite{BKT}.
For the ``{\it solitonic}'' bubbles, another 0 behavior is
the evolution of global monopole itself. Fig. 3 shows that, as far as
the spherical symmetry is kept for the scalar amplitude, the global monopole 
remains
to be stable and its long-range energy tail 
keeps growing before bubble percolation by consuming a part of false
vacuum energy (proportional to the increment of bubble radius) obtained
from the growth of true vacuum bubble (proportional to the increment of
spatial volume).  

\vspace{10mm}
%\input{abfig3}

% GNUPLOT: LaTeX picture
\setlength{\unitlength}{0.240900pt}
\ifx\plotpoint\undefined\newsavebox{\plotpoint}\fi


\vspace{20mm}

In this work we have argued that an appropriate continuous symmetry,
{\it e.g.} global $O(3)$ symmetry, can support a new bubble solution 
for 0-order phase transition. It contains a soliton, {\it e.g.}
a global monopole, from the moment of bubble nucleation and the 
production rate of it can be considerable for a certain shapes of
scalar potentials.

\vspace{5mm}

Y.K.'s research is supported in part by JSPS (No.93033), KOSEF as a
Brain Pool and the Korean Ministry of Education (BSRI-94-2413).

\newpage

\def\hebibliography#1{\begin{center}\subsection*{References
}\end{center}\list
  {[\arabic{enumi}]}{\settowidth\labelwidth{[#1]}
\leftmargin\labelwidth    \advance\leftmargin\labelsep
    \usecounter{enumi}}
    \def\newblock{\hskip .11em plus .33em minus .07em}
    \sloppy\clubpenalty4000\widowpenalty4000
    \sfcode`\.=1000\relax}

\let\endhebibliography=\endlist

\begin{hebibliography}{100}
\bibitem{Lan} J. S. Langer, Ann. Phys. {\bf 41}, 108 (1967).
\bibitem{Col} S. Coleman, Phys. Rev. {\bf D15}, 2929 (1977); C. Callan
and S. Coleman, {\it ibid} {\bf D16}, 1762 (1977).
\bibitem{Aff} I. Affleck, Phys. Rev. Lett. {\bf 46}, 306 (1981);
A. D. Linde, Phys. Lett. {\bf B70}, 306 (1977); {\it ibid} 
{\bf B100}, 37 (1981); Nucl. Phys. {\bf B216}, 421 (1983).
\bibitem{CGM} S. Coleman, V. Glaser and A. Martin, Comm. Math. Phys.
{\bf 58}, 211 (1978); S. Coleman, Nucl. Phys. {\bf B298}, 178 (1988).
\bibitem{CL} S. Coleman and F. De Luccia, Phys. Rev. {\bf D21}, 
3305 (1980).
\bibitem{Kim} Y. Kim, Nagoya University preprint DPNU-94-39, 
hep-th/9410076.
\bibitem{KMS} N. Sakai, Y. Kim and K. Maeda, Nagoya University
preprint, DPNU-94-40; Y. Kim, K. Maeda and N. Sakai, Waseda
University preprint WU-AP/41/94.
\bibitem{KKK} C. Kim, S. Kim and Y. Kim, Phys. Rev. {\bf D47},
5434 (1993).
\bibitem{BV} M. Barriola and A. Vilenkin, Phys. Rev. Lett. {\bf 63}
341 (1989).
\bibitem{HL} D. Harari and C. Loust\'{o}, Phys. Rev. {\bf D42}, 2626
(1990).
\bibitem{Vil} A. Vilenkin, Phys. Rev. Lett. {\bf 72}, 3137 (1994);
A. D. Linde, Phys. Lett. {\bf B327}, 208 (1994);
A. D. Linde and D. Linde, Phys. Rev. {\bf D50}, 2456 (1994).
\bibitem{MSSK} K. Sato, M. Sasaki, H. Kodama and K. Maeda, Prog. Theor.
Phys. {\bf 65}, 1143 (1981); K. Maeda, K. Sato, M. Sasaki and H. Kodama,
Phys. Lett. {\bf B108}, 98 (1982).
\bibitem{Ste} P. J. Steinhardt, Phys. Rev. {\bf D25}, 2074 (1982).
\bibitem{BKT} V. A. Berezin, V. A. Kuzmin and I. I. Tkachev,
Phys. Lett. {\bf B120}, 91 (1983), Phys. Rev. {\bf D36}, 2919 (1987);
S. K. Blau, E. I. Guendelman and A. H. Guth, Phys. Rev. {\bf D35},
1747 (1987).
\end{hebibliography}
\end{document}